\documentclass[a4paper,twocolumn,11pt,accepted=2026-09-15]{quantumarticle}
\pdfoutput=1
\usepackage[utf8]{inputenc}
\usepackage[english]{babel}
\usepackage[T1]{fontenc}
\usepackage{amsmath}
\usepackage{amssymb}
\usepackage{hyperref}
\usepackage[numbers]{natbib}
\usepackage{graphicx}
\usepackage{soul, xcolor}

\usepackage{tikz}
\usepackage{lipsum}

\begin{document}
\setstcolor{red}

\title{Erasing Photon-Number Correlations through Hong-Ou-Mandel Interference}

\author{Fabian Schlue}
\email{fabian.schlue@upb.de}
\affiliation{Integrated Quantum Optics, Institute for Photonic Quantum Systems (PhoQS), Paderborn University,  Paderborn, Germany}
\orcid{0009-0003-0365-7489}
\author{Patrick Folge}
\affiliation{Integrated Quantum Optics, Institute for Photonic Quantum Systems (PhoQS), Paderborn University,  Paderborn, Germany}
\orcid{0009-0008-7178-796X}
\author{Takefumi Nomura} 
\affiliation{Department of Applied Physics, School of Engineering, The University of Tokyo, Tokyo, Japan}
\author{Philip Held} 
\affiliation{Integrated Quantum Optics, Institute for Photonic Quantum Systems (PhoQS), Paderborn University,  Paderborn, Germany}
\author{Federico Pegoraro}
\affiliation{Integrated Quantum Optics, Institute for Photonic Quantum Systems (PhoQS), Paderborn University,  Paderborn, Germany}
\orcid{0000-0003-0535-3756}
\author{Michael Stefszky} 
\affiliation{Integrated Quantum Optics, Institute for Photonic Quantum Systems (PhoQS), Paderborn University,  Paderborn, Germany}
\orcid{0000-0001-5379-3460}
\author{Benjamin Brecht}
\affiliation{Integrated Quantum Optics, Institute for Photonic Quantum Systems (PhoQS), Paderborn University,  Paderborn, Germany}
\orcid{0000-0003-4140-0556}
\author{Stephen M. Barnett}
\affiliation{School of Physics and Astronomy, University of Glasgow, Glasgow G12 8QQ, UK}
\orcid{0000-0003-0733-4524}
\author{Christine Silberhorn}
\affiliation{Integrated Quantum Optics, Institute for Photonic Quantum Systems (PhoQS), Paderborn University,  Paderborn, Germany}
\orcid{0000-0002-2349-5443}
\maketitle

\begin{abstract}
  A parametric down-conversion source interfering on a beam splitter can be described as both a source of entangled N00N-states or a source of independent, and thus uncorrelated squeezers. The disparity between these outcomes can be attributed to whether one takes a standard discrete- or continuous-variable approach to describing the system. More precisely, this difference in output is due to the types of measurements involved and the way in which the data is processed, as both setups are identical, clearly identical states are produced. Here we take a hybrid approach to describe the state, which is generated by parametric down-conversion as input state, and investigate the measurements of the output ports of the interferometer using photon number resolved detection. We show that the output of this interference is separable in the photon number picture and demonstrate the worth of photon-number correlation measures for the characterization of squeezed light sources for hybrid applications like Gaussian boson-sampling.
\end{abstract}

\section{Introduction}
Photonics is an appealing platform for the implementation of numerous quantum technologies, such as quantum metrology and sensing, quantum communication, quantum simulation, and quantum computation. Traditionally, due to the distinct wave-particle duality of optical fields, quantum photonics has been conceptually approached from two directions: on the one hand, continuous variable quantum optics (CVQO) focuses on the field aspect of light and investigates field quadratures to reveal effects such as squeezing; while on the other hand, discrete variable quantum optics (DVQO) employs the particle aspect of light through photon counting methods. Although DVQO and CVQO have traditionally been considered independently, new hybrid approaches, ones that combine the tenets of CVQO and DVQO, provide a more complete understanding and allow us to discover new effects.

It has been demonstrated that these hybrid investigations are capable of revealing new effects and physics, providing an opportunity to further develop our understanding of quantum systems \cite{Andersen2015}. One high-impact example of such a hybrid system is Gaussian Boson Sampling (GBS), which combines both approaches by interfering squeezed states in a large Haar-random beam splitter network with photon counting at the output of the network \cite{Hamilton2017,Kruse2019}. This hybrid approach  was used to demonstrate quantum advantage. \cite{Zhong2020,Madsen2022}.

Although large hybrid networks such as those used in GBS can provide extremely powerful and interesting platforms for testing fundamental physics, interesting physics can already be found at the smallest possible network scale, i.e. a single beam splitter with two inputs and two outputs. One prominent source of input light to this beam splitter is parametric down-conversion (PDC). The interference of PDC light on the balanced beam splitter is often considered in different ways in the two areas of quantum optics. In CVQO the PDC output is considered a two-mode squeezed vacuum state (TMSVS) \cite{Eckstein2011}, an entangled state, also known as EPR state \cite{Einstein1935}. Inputting this state into the two inputs of the beam splitter results in two separable single-mode squeezed vacuum state (SMSVS) \cite{Barnett2005}. The opposite direction (the interference of two SMSVS on a balanced beam splitter) is often used to create TMSVS \cite{Takeda2019,Larsen2019}. For TMSVS the presence of non-classical correlations between the two beams of light is typically measured using the difference current of two photodetectors, also known as intensity difference squeezing \cite{Zhang2002,Bondani2007}. The measured intensity correlations of the state are proportional to the photon numbers that are present, but the measurement is not resolving them. Therefore, this method is missing the photon number information of the state

In the DVQO or low gain regime the PDC source is often approximated as a source of post-selected indistinguishable photon pairs. Taking indistinguishable photon pairs as an input to the beam splitter results in the Hong-Ou-Mandel interference (HOMI) experiment \cite{Hong1987}, where the interference visibility is an important benchmark for the quality of the indistinguishable photons. The interference output is a maximally entangled N00N-state \cite{Dowling2008}. These measurements are typically performed using coincidence detection between two click detectors. Click detectors can resolve the presence or absence of photons, but not the exact number of them. This measurement truncates the state to the two-photon subspace. It also conditions the state on the presence of at least one photon, or in other words removing the vacuum component, implicitly performing the aforementioned post-selection of the PDC state. This conditioning of the photon statistics is creating the entanglement observed in the N00N-state \cite{Marcovitch2007, Tasca2009, Blasiak2021}.

However, the two descriptions must be equivalent, and this must also hold true independent of the gain level. Furthermore, in the absence of post-selection, one would expect the output state to also be decorrelated in the photon number picture, and not only in the intensities. Here we experimentally verify for the first time that the photon number statistics generated by HOMI of a TMSVS are in fact decorrelated by measuring the statistics of the two output modes up to eight clicks. We do this by evaluating the present correlations with three correlation measures applied to the joint photon number distributions. Our simulations, containing no free parameters and assuming a single spectral mode TMSVS, show very good agreement with our results, a testimony to our high quality single spectral mode source. We also note that with access to the complete joint photon number statistics of this interference one can deepen the understanding of the effects of data analysis on entanglement interpretations between DVQO and CVQO. This together with the importance of considering higher-order photon number components, hopefully elevates the progress in hybrid quantum optics systems in the future. 

In addition, we demonstrate that our method can directly assess the quality of the SMSVSs generated through the interference of a TMSVS, specifically by evaluating their separability. This is an important metric in the context of high-dimensional photonic networks, e.g. Gaussian boson sampling, which requires single spectral-mode SMSVS as a resource.

\section{Concept and Theory}
\label{sec:concept}
\begin{figure}
	\includegraphics[width=\linewidth]{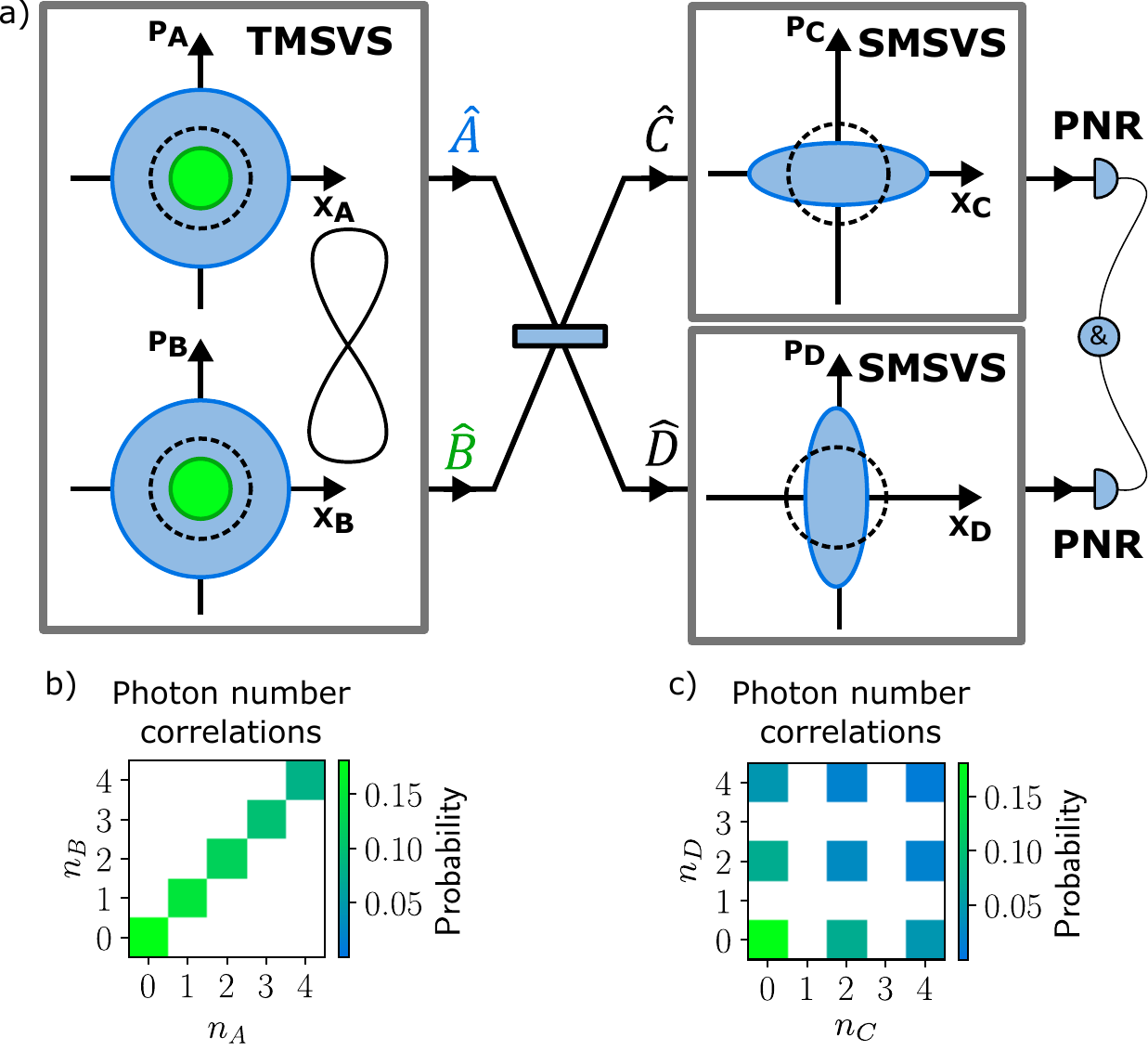}
	\caption{\label{fig:concept}Conceptual representation of the presented experiment. a) A TMSVS is interfered on a 50:50 beam splitter, resulting in two SMSVSs, $\text{P}_i$ and $\text{X}_i$ are field quadrature amplitudes in the respective beam splitter ports. The resulting photon number statistics are measured with a photon number resolving (PNR) detector. b) and c) Joint photon number distributions for a TMSVS and two SMSVSs, respectively, $n_i$ is photon number in the respective output.}
\end{figure}

The concept of this work is shown in FIG.~\ref{fig:concept}. We begin with the phase space representation of a quadrature entangled TMSVS (FIG.~\ref{fig:concept} a), left). Here $\text{P}_i$ and $\text{X}_i$ for $i \in \{A,B\}$ correspond to the field quadrature amplitudes. The state is characterized by each individual mode having larger quadrature fluctuations $\langle(\Delta\hat{X}_i)^2\rangle$ and $\langle(\Delta\hat{P}_i)^2\rangle$ (blue area) than the vacuum (black dotted circle). However, the measurement of one subsystem allows us to infer the properties of the second subsystem to below the quantum limit (green area), i.e. reduced conditional variance, or EPR entanglement \cite{Reid2009}. 

Interfering the TMSVS on a 50:50 beam splitter will result in two independent, phase shifted, SMSVS \cite{Bouwmeester2001}, see FIG.~\ref{fig:concept} a), right panels. SMSVSs are characterized by exhibiting quadrature fluctuations below the vacuum level along one axis (squeezing) and fluctuations higher than the vacuum level along the orthogonal axis  (anti-squeezing).

The simulated joint photon number distributions $P(n_C,n_D)$ for lossless TMSVS and two SMSVSs are shown in FIG.~\ref{fig:concept} b) and c). For more information see Appendix \ref{a:jointprob}. 
In FIG.~\ref{fig:concept} b) one can clearly see perfect photon number correlations of the TMSVS. However, in FIG.~\ref{fig:concept} c) although it may not be obvious, the resulting joint photon number distribution is uncorrelated. The distinction between correlated or uncorrelated distributions becomes even less obvious in a realistic scenario including noise and loss. 


To measure the amount of correlation we first introduce the correlation coefficient corr which is the normalized form of the covariance $\text{cov}$ \cite{Asuero2006}
\begin{equation}
	\text{corr}(P_C,P_D) = \frac{\text{cov}(P_C,P_D)}{\sigma_{P_C} \sigma_{P_D}},
	\label{eq:corr}
\end{equation}
where $P_C$ and $P_D$ are the two marginal photon number distributions after the beam splitter and $\sigma_{P_C(P_D)}$ are their respective standard deviations. The correlation coefficient is bounded between 1 and -1, indicating perfect correlations and anti-correlations respectively. When the marginal photon number distributions $P_C$ and $P_D$ are completely uncorrelated the correlation coefficient is zero. Note that the correlation coefficient in Eq.~\eqref{eq:corr} is sensitive only to the degree of second-order correlation \cite{Asuero2006}, therefore, a state exhibiting higher-order correlations can have a correlation coefficient of 0. In the framework of CVQO it is common to focus on so-called Gaussian states, states that exhibit a Gaussian Wigner function,such as the TMSVS and the SMSVS. These are fully characterized by their first and second order moments, hence the correlation coefficient is well suited for evaluating properties of these states. 

As a second method, we will utilize the Schmidt decomposition. In contrast to the correlation coefficient, this method is sensitive to higher-order correlations. In \cite{Ekert1995}, the Schmidt decomposition for any pure quantum state $|\psi\rangle \in \mathcal{H} = \mathcal{H}_u \otimes \mathcal{H}_v$, which is a state of a composite system, is given by
\begin{equation}
	|\psi\rangle = \sum_{i=0}^{\infty}g_i|u_i\rangle\otimes|v_i\rangle.
	\label{eq:schmidt_decomposition}
\end{equation}
Here, the $\{|u_i\rangle\}$ and $\{|v_i\rangle\}$ form orthonormal bases of $\mathcal{H}_u$ and $\mathcal{H}_v$, respectively, and $g_i\in\mathbb{C}$. In particular, it is shown that a TMSVS is decomposed as
\begin{equation}
	|\psi\rangle_\mathrm{TMSVS} = \hat{S}_{ij}(\xi)|0\rangle = \sum_{n=0}^{\infty}c_n|n\rangle_u\otimes|n\rangle_v,
	\label{eq:photon_number_decomposition}
\end{equation}
where $\hat{S}_{ij}(\xi)$ is the two-mode squeezing operator in spatial modes $i$ and $j$ with squeezing parameter $\xi\in\mathbb{C}$ (see Eq.~\eqref{eq:squeez}), the $c_n$ are the complex amplitude coefficients, and the $|n\rangle$ are photon-number Fock states. 

The Schmidt decomposition often appears in the context of photon-pair states generated in parametric down-conversion \cite{Law2000}. Here, we decompose the so-called joint spectral amplitude $f(\omega_s, \omega_i)$ that describes the spectral content of the PDC state as $f(\omega_s, \omega_i)=\sum_{i=0}^\infty\sqrt{\lambda_i}\psi_i(\omega_s)\phi_i(\omega_i)$, with the $\psi_i(\omega_s)$ and $\phi_i(\omega_i)$ describing the spectral modes \cite{Brecht2015} of signal and idler, respectively. The Schmidt decomposition yields a measure for the modal content of the photon-pair state, the so-called Schmidt number $K_f = 1/\left(\sum_{i=0}^{\infty}\lambda_i^2\right)$, where the subscript $f$ denotes the spectral mode basis of the decomposition. This number provides the effective number of spectral modes that the state occupies and simultaneously quantifies the spectral entanglement between the two photons. It ranges from $K_f=1$ for a separable, spectrally single-mode state to $K_f\rightarrow\infty$ for a perfectly correlated and hence spectrally maximally entangled state. 

In this work, we apply the notion of the Schmidt number to the photon number basis (see Eq. \eqref{eq:photon_number_decomposition}), where the modes $u$ and $v$ now label the two output ports of the beam splitter $C$ and $D$. We assume that the state to which we apply the Schmidt decomposition is spectrally single mode\textemdash a condition that is approximately true for the TMSVS resource generated by our source ($K_f=1.3$)\textemdash where the Schmidt number $K_p$ is now a measure for the photon-number correlation between the two subsystems $\mathcal{H}_u$ and $\mathcal{H}_v$. The subscript $p$ here indicates the we interpret the Schmidt number in the photon number picture. We note that we apply the Schmidt decomposition to the measured intensities instead of the amplitudes of the states. This is justified because phases between different photon-number components would not change the decomposition. 

The third method we use is the mutual information $I(P_C,P_D)$ \cite{Cover2006}. This measure, like the Schmidt number, is sensitive to higher-order correlations and is often used in information sciences. It is defined as
\begin{equation}
	I(P_C,P_D) = \sum_{n_c,n_d} P(n_c,n_d) \log_{10} \frac{P(n_c,n_d)}{P(n_c)P(n_d)},
\end{equation}
where $P(n_c,n_d)$ is the joint photon number distribution and $P(i)$ is the respective marginal distribution of $P(n_c,n_d)$. The mutual information is always $I(P_C,P_D)\geq 0$ and $I(P_C,P_D)=0$ if and only if $P(n_c,n_d) = P(n_c)P(n_d)$ is separable.

\section{\label{sec:exp}Experiment}
\begin{figure}
	\includegraphics[width=\linewidth]{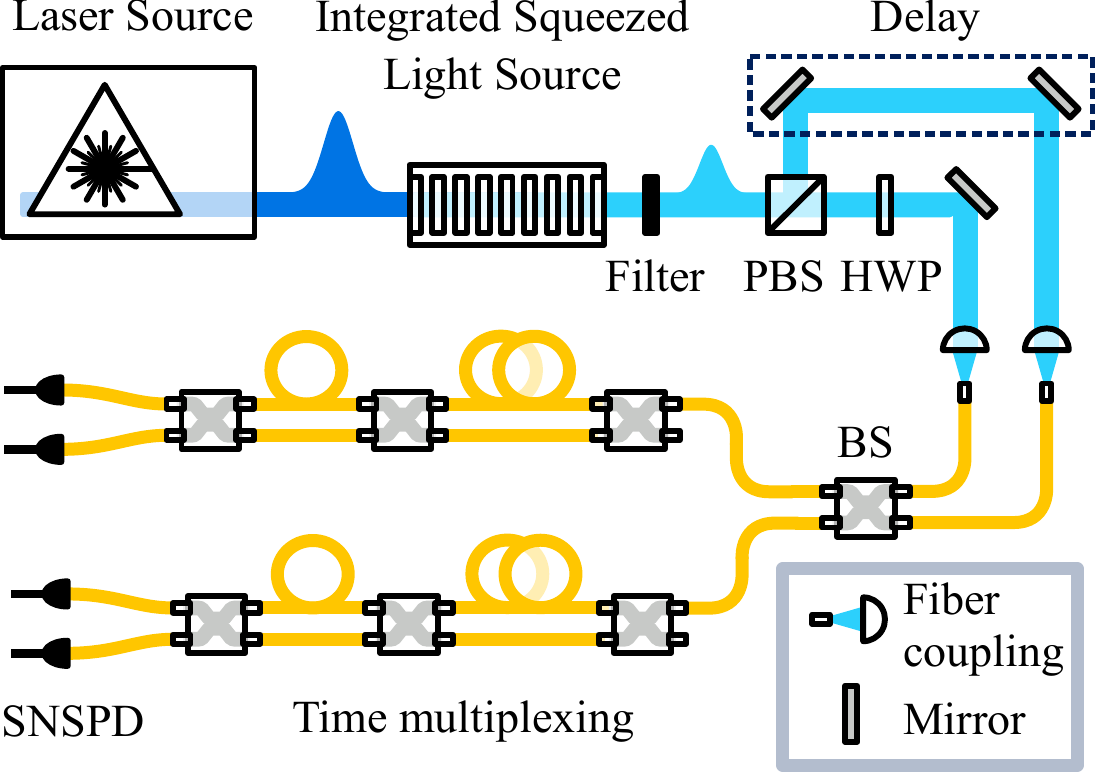}
	\caption{\label{fig:setup}Schematic representation of the experimental setup. An integrated squeezed light source is pumped by a pulsed laser. The resulting TMSVS is filtered and split on a polarizing beam splitter (PBS). The polarization of one output of the PBS is rotated by a half-wave plate (HWP) to interfere both outputs on a balanced fiber beam splitter (BS) with an adjustable delay. The click statistics are measured via time-multiplexed detection on superconducting nano-wire single photon detectors (SNSPD).}
\end{figure}

We generate our TMSVS with a type-II PDC source in a 25mm long periodically poled potassium titanyl phosphate waveguide \cite{Pegoraro2023} (see FIG.~\ref{fig:setup}). Due to careful dispersion engineering, the source generates nearly a single spectral-mode, pulsed, spectrally degenerate TMSVS. We find an effective number of spectral-modes of $K_f=1.3$. For this we make sure that the pump pulse spectral width is matched to the phase matching bandwidth \cite{Eckstein2011,Harder2016}. We verify this using the second order normalized correlation function measure $g^{(2)}(0)$ and achieved a value of $1.75 \pm 0.05$, which can be used to determine the effective mode number \cite{Christ2011}. Additionally, we verified a high interference visibility of the created photons in a HOMI experiment and reached $94\%$ visibility, highlighting the photons indistinguishability. In this work, this requires pump pulses with a duration of 3 ps and a central wavelength of 772.5 nm, which we derive from an ultrafast oscillator with a repetition rate of 76.4 MHz. We reduce the pulse repetition rate to 200 kHz using a pulse picker to facilitate time-multiplexed photon counting \cite{Avenhaus2010}. After the source we use a wavelength filter to suppress the pump light and a 1.8~nm wide spectral band-pass filter that is matched to the central peak in the generated spectrum. The band-pass filter removes phase matching sidebands which would increase the number of spectral modes \cite{MeyerScott2017}, while giving minimal attenuation to the central peak. It is used in all presented measurements.

The generated photon pairs are split on a polarizing beam splitter (PBS) and interfere at the following balanced fiber beam splitter (BS). One output of the PBS possesses an adjustable delay, enabling variable temporal overlap between the two outputs of the PBS at the fiber BS. In the other output of the PBS a half wave plate (HWP) rotates the polarization of the created photons by $90^\circ$ to ensure polarization indistinguishability at the fiber BS.

The photon number resolution (PNR) in our experiment is achieved via two independent 8 time-bin time-multiplexed detectors (TMDs) \cite{Achilles2003}. The first TMD, hereafter referred to as the high efficiency arm, has a time-bin separation of 100~ns, compatible with a superconducting nano-wire single photon detector (SNSPD) system with 70~ns dead time and $>90\%$ quantum efficiency. The second is optimized for speed and has a time-bin separation of 17~ns, compatible with an SNSPD system with 13~ns dead time and $>70\%$ quantum efficiency, hereafter referred to as the low efficiency arm. One could have used a single TMD for the measurement of the joint photon number distributions, however, our devices lack a long fiber delay on the second input required to use both inputs. This results in slightly asymmetric efficiencies for the two measured outputs of the PDC. The total efficiencies are $20\% \pm 6\%$ and $14\% \pm 6\%$ for the high and low efficiency arm respectively. For more detail see Appendix \ref{a:losses}. The signals from the 4 SNSPDs are recorded using a time to digital converter. By calibrating the temporal delay between the reference laser trigger and the individual time bins of both TMDs we are able to narrowly time-filter the bins within a 2~ns window, thereby reducing noise counts. With this calibration we correlate the number of detection events in each output on a shot-by-shot basis to measure the joint click distribution of the input state with up to 8 detection events per output \cite{Tiedau2020}. 

We have measured temporal delays ranging from around -7~ps to 2~ps, where 0~ps corresponds to perfect temporal overlap between the input pulses. The HOMI is expected to be symmetric around 0~ps, we therefore have taken a few measurements above 0~ps to confirm this. Each measured joint photon number distribution contains 228 million experimental runs. We describe in the next section how we convert the measured data to the joint photon number distribution. 

Obtaining high quality data requires one to carefully choose the operating power. We operated the source at a measured mean photon number of $0.060 \pm 0.008$ in the high efficiency arm. This corresponds to having $P(1,1) = 6\cdot 10^{-3}$, $P(2,2) = 9\cdot 10^{-5}$ and $P(3,3) = 1\cdot 10^{-6}$. This low mean photon number is far away from detector saturation effects.  Note that we are operating in a regime where the DVQO approximation of single photon pair emission, neglecting higher photon number components, is still often assumed \cite{Jin2015,Kobayashi2016}. However, we are explicitly measuring them and show their impact here.


\section{Results}
The joint click statistics of the interference of a TMSVS at a BS with varying temporal overlap between the inputs was measured. First, the joint click statistics need to be converted to pseudo photon number correlations, due to the probabilistic nature of splitting the input pulse inside the TMD. There are ways of correcting for this so-called convolution \cite{Achilles2004,Krishnaswamy2024}, by using the fact that we know how likely it is to get multiple photons in one detection bin, assuming equal splitting in the TMD. To recover the photon number statistics, we use the method presented in \cite{Krishnaswamy2024}, which is applied to all data before proceeding with further analysis.  

\begin{figure*}
	\includegraphics[width=\linewidth]{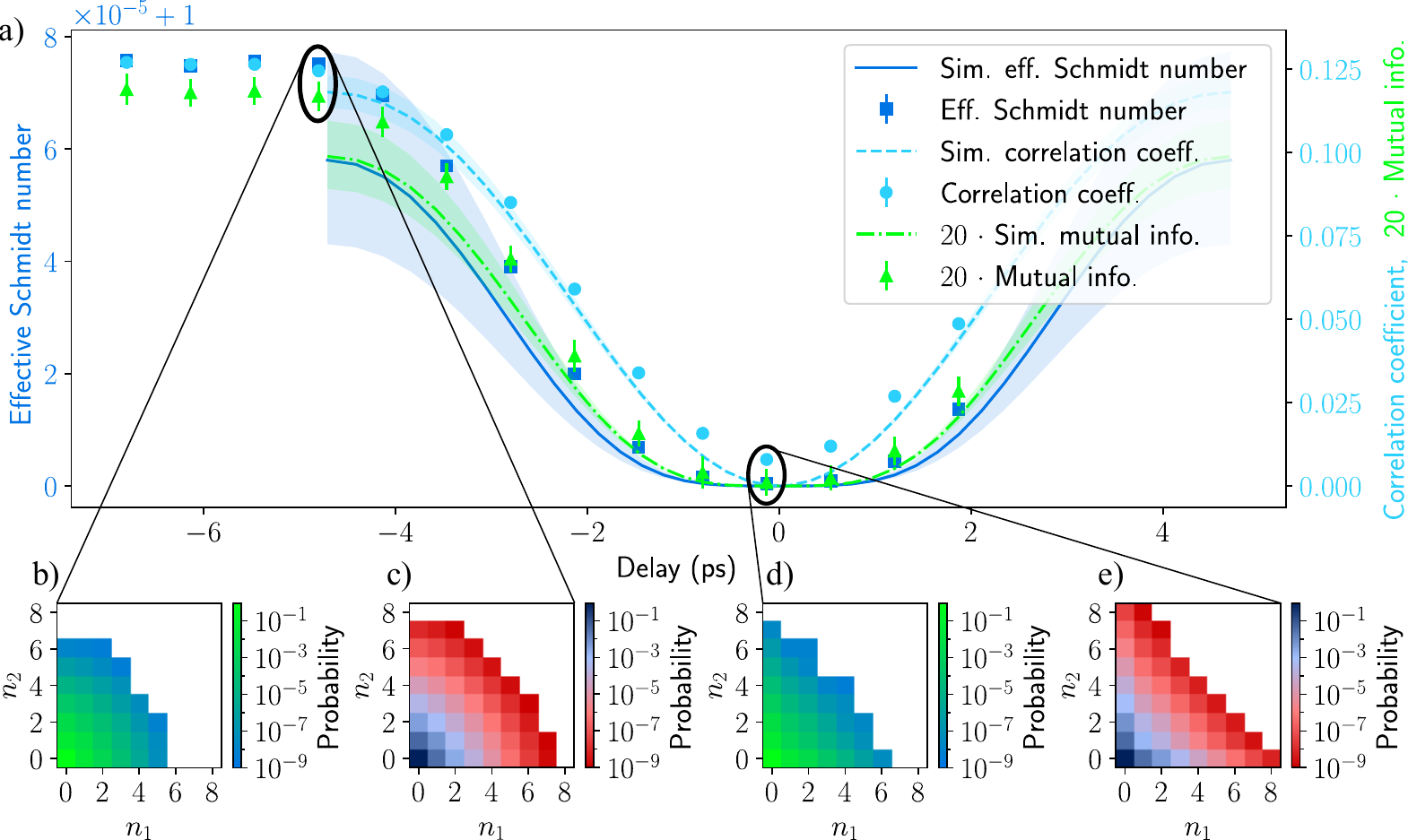}
	\caption{\label{fig:effk}a) Correlation measures as the temporal delay between the two pulses is varied. Lines show simulations for measured mean photon number. Shaded areas show simulations performed for upper and lower standard deviation from the measured mean photon number. b) and d) measured joint photon number distribution for high and zero temporal delay as indicated in a), $n_i$ photon number. c) and e) simulated joint photon number distribution corresponding to the same temporal overlaps as the measured distributions.}
\end{figure*}

We compare our experimental results to simulations detailed in Appendix \ref{a:sim}. The simulation takes the measured squeezing strength and losses as input parameters. The squeezing strength $r$ for a single spectral mode TMSVS is back-calculated from the measured mean photon number $\bar n$ and losses $\eta$ in the experiment via $r=\text{arsinh}(\sqrt{\bar n /\eta})$. The mean photon number has fluctuated during the experiment, which resulted in it being the largest experimental uncertainty. We have observed that the simulations are most sensitive to variations in the mean photon number. We therefore perform the simulations for three values, the measured mean photon number and one standard deviation below and above it. The lines in Fig. \ref{fig:effk} correspond to the measured mean photon number. The shaded areas in Fig. \ref{fig:effk} are the values between the simulation for the upper and lower standard deviation from the measured mean. The size of the shaded areas illustrates the sensitivity of the simulation to the mean photon number and therefore also indicates possible deviations of the simulation and the measured values due to mean photon number fluctuations in the experiment. We used the measured losses from the experiment and distribute the losses before and after the interference at the 50:50 beam splitter as measured in the experiment (see Appendix \ref{a:losses}). From this we obtain the joint photon number distribution and proceed with the same analysis as with the experimental data.  

The results of our analysis can be seen in FIG.~\ref{fig:effk}. In FIG.~\ref{fig:effk} a) the measured values for the three correlation measures and the results from their respective simulations are shown. In general, one sees that all three correlation measures follow their respective simulations rather closely. We compare our measured data to our theory with no free parameters based on an ideal TMSVS at the input of the beam splitter and only account for experimentally measured efficiencies. The good agreement between experiment and simulation is testimony to the high single mode quality of our source. 

There are, however, some deviations of the measured values from the simulations. One deviation can be seen for temporal separations greater than $-4$~ps, where the experimental values are on the upper edges of the confidence intervals of the simulation. In appendix \ref{a:multi_sim} we show preliminary simulations including approximations of spectral multimode behavior, which indicate that the overlap between simulation and measurement could be improved when multimodeness is considered. However, a more rigorous simulation approach, e.g. following \cite{Bulmer2026}, is required to make more conclusive statements.

Another deviation is present for the correlation coefficient at zero time delay where the simulation and measurement do not overlap. With the parameters of our current simulation we are unable to recreate a non-zero correlation coefficient at zero time delay, this includes the preliminary multimode simulations. Further investigations are needed to determine the origin of this effect.

All three correlation measures in FIG.~\ref{fig:effk} a) show correlations for temporal delays larger than $-4$~ps. This is consistent with the fact that at this temporal separation, the TMSVS pulses, with duration of $3$~ps, do not exhibit a significant temporal overlap. Therefore, minimal interference takes place. In contrast, for a temporal delay of $0$~ps, the value of the Schmidt number is very close to 1 and the mutual information is consistent with 0 inside the error margin, due to near perfect interference. This indicates that the outputs are very close to independent, as one would expect in the ideal case of two independent squeezers. The correlation coefficient shows a similar shape as the other two measures but does not go to 0.

Another interesting property of the correlation coefficient, seen in both theory and measurements, is its higher sensitivity to pulse overlap of the input pulses, as the distributions are narrower (see FIG.~\ref{fig:effk} a)). This indicates that the correlation coefficient can be a better measure for finding the best temporal overlap in the measured joint photon number distributions \cite{Lyons2018}. 

Next, we investigate the joint photon number distributions themselves, illustrated in FIG.~\ref{fig:effk} b) and d). These correspond to the regions indicated in FIG.~\ref{fig:effk} a). One can see that the measured joint photon number distributions are substantially different to the simulated lossless cases shown in FIG.~\ref{fig:concept} b) and c). This is due to the losses in the experiment eliminating the strict photon number correlations of the quantum state. The simulated joint photon number distributions, including experimental losses and matched to the experimental settings of FIG.~\ref{fig:effk} b) and d), are shown in FIG.~\ref{fig:effk} c) and e), confirming that losses are the main cause for the difference to the ideal case in FIG.~\ref{fig:concept} b) and c).

\begin{figure}
	\includegraphics[width=\linewidth]{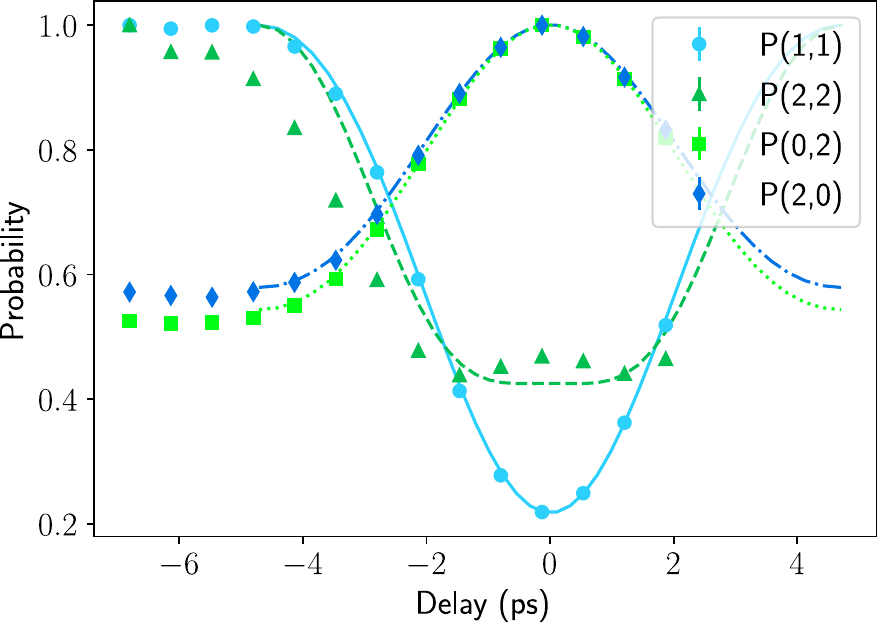}
	\caption{\label{fig:pncomp}Probability of individual photon number components normalized to maximum occurrence. Measured values from the joint photon number distributions, together with the respective probabilities from simulation.}
\end{figure}

In FIG.~\ref{fig:pncomp} we show four specific photon number components from the joint photon number distributions over the complete range of temporal delays measured. First, we show the standard DVQO HOMI dip, the probability of having one photon in each output $P(1,1)$. This shows very good agreement with our simulation. Second, we show the probability of having two photons in each output $P(2,2)$. Previous work has shown that there is an increase in probability at zero time delay \cite{Ferreri2019} for this output pattern. In the measured $P(2,2)$ probability we see the expected increase of probability at zero time delay, which is not present in our single mode simulation. However, the presence of multiple modes in the experiment increases the probability of having a $P(2,2)$ event at zero time delay \cite{Ferreri2019}, which we confirm with our multimode simulations (see appendix \ref{a:multi_sim}). Lastly, the probabilities of having two photons in either output and zero in the other $P(2,0)$ and $P(0,2)$ are presented. They are also often referred to as bunching peaks. One can clearly see a significant increase in the occurrence of these two components towards perfect interference (zero time delay). This increase, together with the decrease in the $P(1,1)$ component, indicates the creation of N00N-states at zero time delay \cite{Dowling2008}. 

\begin{figure*}
	\includegraphics[width=\linewidth]{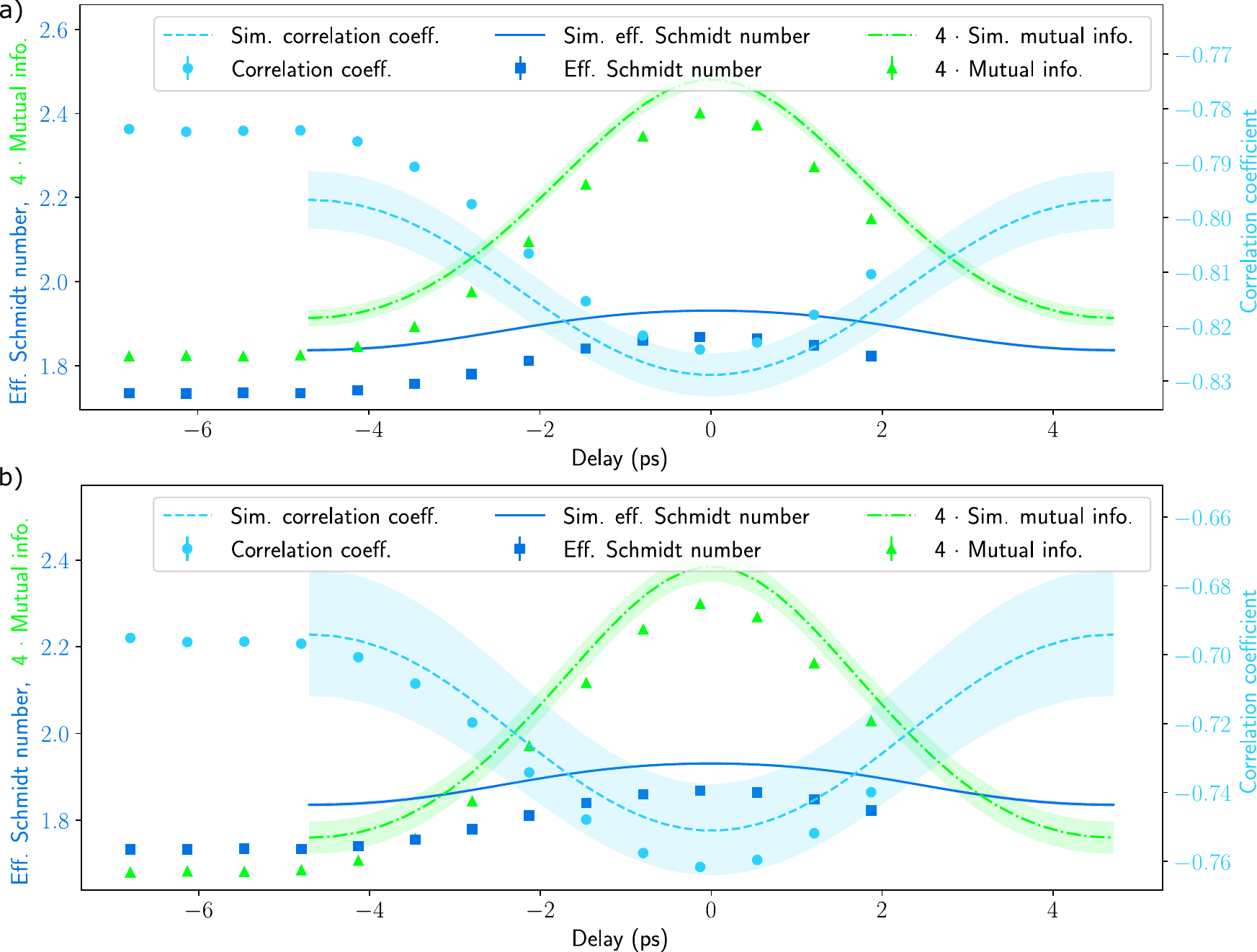}
	\caption{\label{fig:trunc}Correlation measurements at different temporal overlaps of the input pulses, a) two-photon subspace with removed vacuum, b) only vacuum-component removed.}
\end{figure*}

One can more readily see the creation of the N00N-state by repeating the previous correlation analysis with a truncated joint probability distribution (see FIG.~\ref{fig:trunc}). For the truncation in FIG.~\ref{fig:trunc} a) we remove all photon number components containing more than $2$ photons and condition on the detection of at least one photon, removing the vacuum component. This describes the typical DVQO HOMI experiment. Note that the removal of higher-order photon number components is not necessary to observe the described behavior, only the removal of the vacuum component is necessary. Removing the higher-order photon number components does not change the outcome of this analysis significantly (see FIG.~\ref{fig:trunc} b)). 

All three correlation measures in FIG.~\ref{fig:trunc} show increased correlation at zero temporal delay compared to the values at large temporal delay and good agreement between simulation and measurement. Note that the correlation coefficient is negative and therefore possesses a dip at zero time delay. Although there is an offset between the measured values and our simulation, they follow the same trend. We see in our simulation that this offset is sensitive to the squeezing parameter, losses, and the multimode contribution, however, as these values are measured, we keep them and accept the offset here. We see in our preliminary multimode simulations (see appendix \ref{a:multi_sim}) that the addition of a second TMSVS reduces the offset for large temporal offsets, but at zero temporal delay the offset remains. Further investigations with a more sophisticated multimode treatment, e.g. \cite{Bulmer2026}, are required to make better conclusions of the offset's origin.

\section{Discussion}
The presented hybrid interference experiment combines the TMSVS input, the state produced by PDC, with a grater than two photon photon-number resolved measurement for the first time \cite{Ferreri2019}. From textbooks we know that CVQO predicts that this interference experiment creates a separable state. As DVQO and CVQO perform the same experiment the total state in the DVQO picture has to also be separable. Therefore, the complete joint photon number statistics has to be uncorrelated, which we have shown experimentally for the first time. The key difference between the field measurements in CVQO and the photon counting measurements in typical DVQO HOMI is the conditioning of the photon counting measurements on having \textit{at least one photon}, effectively removing the vacuum component $P(0,0)$ from the joint probability distribution. This conditioning introduces correlated information about the two-mode state, which leads to the creation of an entangled N00N-state, see Appendix \ref{a:trunc} for a more detailed explanation. It follows that the measurement of higher-order photon number components will not make the state separable when the $P(0,0)$ probability is removed, as confirmed in our experiment. 

In literature the interference on a balanced beam splitter is well explored in the two points of perfect temporal overlap and very large temporal offset between the pulses. In these cases, the interference can be explained in a single mode picture in every input and output mode. However, when performing a complete temporal scan of the two inputs, as done here, this treatment as a single mode is no longer possible. For the presented case at least four modes are needed to simulate the behavior (see Appendix \ref{a:sim}). Therefore, correlations in the region of partial overlap of the input pulses might be interesting for further investigations. For this the presented experiment, including higher-order photon number components, might prove to be a valuable test platform.

Another application for our presented experiment might be in quantum metrology, where the interference of the $|n,n\rangle$ state on a balanced beam splitter is used to produce Holland-Burnett states \cite{Holland1993}. These states approximate the interferometric qualities of N00N-states, but are almost optimally resilient to experimental imperfections \cite{Datta2011}. By slight adaptations of our experiment, metrology applications similar to the ones shown in literature are possible \cite{Xiang2013,Thekkadath2020}. Crucial in the context of this work is that in order to observe the Holland-Burnett states the measured results are post selected on the presence of the desired photon number combinations.

As a final note, we have shown that photon number resolved correlation measurement paired with the three types of correlation measures enables us to accurately measure the amount of residual correlation after the transformation from TMSVS to SMSVS. With this we have a direct measure of the remaining correlations, improving upon the current practice of inferring the quality of the SMSVS by characterizing the TMSVS. Therefore, this method presents a new characterization technique for SMSVS that is particularly relevant when generating them by interfering the two spatial modes of a TMSVS.

\section{Conclusion}
In this study we have investigated the interference of a TMSVS on a single balanced beam splitter in the photon number picture. Specifically, we performed a Hong-Ou-Mandel-type experiment by interfering a close to perfectly single spectral mode TMSVS generated by a type-II PDC source on a balanced beam splitter and measuring the photon number correlations at different temporal overlaps of the two input modes from the TMSVS. Using the more complete CVQO approach that considers the whole field, we know that for perfect temporal overlap the resulting state is separable. We show for the first time that the complete joint photon number statistics of this state is indeed separable. We paired this measurement with an idealized simulation that contains no free parameters, only measured losses and mean photon number as parameters and assumes a perfect single spectral mode TMSVS as input. The good agreement between our experiment and simulation highlight the high quality close to single spectral mode character of our source. We also note the implicit difference between the typical CVQO and DVQO interpretations of this experiment explicitly, CVQO considers the complete state including the vacuum component, DVQO conditions on the existence of at least one photon, thereby removing the vacuum component.  We propose this photon number correlation measurement for the characterization of SMSVS as a direct measure of remaining correlation in the desired single spectral mode SMSVS created from type-II PDC, advancing the development of sources for hybrid applications. With slight modifications of our experiment also metrology applications in the framework of Holland-Burnett states are possible. Besides these immediate applications the presented experiment also sparks interest in further investigating the correlation structure in the regions of partial overlap of the input fields, as there the description with a single mode theory is not sufficient anymore. 

\begin{acknowledgments}
	F.S. is part of the Max Planck School of Photonics supported by the Dieter Schwarz Foundation, the German Federal Ministry of Research, Technology and Space (BMFTR), and the Max Planck Society. SMB is supported by the Royal Society (RSRP/R/210005). This work has received funding from the BMFTR within the PhoQuant project (Grant No. 13N16103) and from the European Commission through the Horizon Europe project EPIQUE (Grant No. 101135288).
\end{acknowledgments}

\section*{Author Contributions}
FS performed the experiment with support from TN, PH and FP. PF and FS performed the simulation with support from TN. FS wrote the manuscript. SB contributed the conditioning effects section. CS, MS and BB directed the research. All authors edited the paper.

\appendix

\section{\label{a:jointprob}Joint Probability Distributions}
Here we describe how the joint probability distributions $P_{AB}(n_1,n_2)$ and $P_{CD}(n_1,n_2)$ from FIG.~1 b) and c) in the main text are obtained. We start from a TMSVS $\left| \psi \right\rangle_{\text{TMSVS}}$ \cite{Barnett2005}
\begin{equation}
	\left| \psi \right\rangle_{\text{TMSVS}} = S_{AB}(\xi)\left| 0\right\rangle_A \left|0 \right\rangle_B
\end{equation}
where $S_{AB}(\xi)$ is the two-mode squeezing operator (see Eq. \eqref{eq:squeez}) and $\left| 0\right\rangle_{A(B)}$ is the vacuum state in modes $A$ and $B$. In the photon number representation this state can be written as
\begin{widetext}
\begin{equation}
	\left| \psi \right\rangle_{\text{TMSVS}} = \frac{1}{\cosh r}\sum_{n=0}^\infty (-1)^n e^{in\phi} (\tanh r)^n \left| n\right\rangle_A \left|n \right\rangle_B.
\end{equation}
The joint photon number distribution $P_{AB}(n_1,n_2)$ for this state is 
\begin{align}
	P_{AB}(n_1,n_2) & = \left| {}_A\left\langle n_1 \right| {}_B\left\langle n_2 \right| \psi \rangle_{\text{TMSVS}}\right|^2\notag\\
	&=\begin{cases}
		(\cosh r)^{-2}(\tanh r)^{2n},  & \text{for } n_1 = n_2,\\
		0, & \text{for }n_1 \neq n_2.
	\end{cases} 
\end{align}
This shows the perfect photon number correlation present in FIG.~1 b) in the main text.

It can be shown that the interference of a TMSVS on a balanced beam splitter results in 
\begin{equation}
	\left| \psi \right\rangle_{CD} = \left| \psi \right\rangle_{\text{SMSVS, }C} \otimes \left| \psi \right\rangle_{\text{SMSVS, }D} 
\end{equation}
with 
\begin{align}
	\left| \psi \right\rangle_{\text{SMSVS}} & = \exp\left[\frac{1}{2}(\xi^*\hat a^2 - \xi \hat a^{\dag 2})\right] \left| 0 \right\rangle \notag \\
	&= \frac{1}{\sqrt{\cosh r}} \sum_{n=0}^\infty(-1)^n \frac{\sqrt{(2n)!}}{2^n n!} e^{in\phi} (\tanh r)^n \left| 2n \right\rangle. 
\end{align}
The photon number probability for this state is given by
\begin{align}
	P(n)&= \left| \left\langle n \right| \psi \rangle_{\text{SMSVS}}\right|^2 \notag\\
	&=\begin{cases}
		\frac{(2n)!}{2^{2n} (n!)^2} \frac{(\tanh r)^{2n}}{\cosh r},  & \text{for } n \text{ even},\\
		0, & \text{for }n \text{ odd}.
	\end{cases}
\end{align}
The joint probability distribution of the state $\left| \psi \right\rangle_{CD}$ can than be written as
\begin{align}
	P_{CD}(n_1,n_2) & = \left| {}_C\left\langle n_1 \right| {}_D\left\langle n_2 \right| \psi \rangle_{\text{SMSVS, }C} \left| \psi \right\rangle_{\text{SMSVS, }D}\right|^2\notag\\
	& = \left| {}_C\left\langle n_1 \right|  \psi \rangle_{\text{SMSVS, }C} \cdot {}_D\langle n_2\left| \psi \right\rangle_{\text{SMSVS, }D}\right|^2\notag\\
	& = \left|{}_C\langle n_1 \left|  \psi \right\rangle_{\text{SMSVS, }C} \right|^2 \cdot \left|{}_D\langle n_2\left| \psi \right\rangle_{\text{SMSVS, }D}\right|^2 \notag\\
	&= P_C(n_1) P_D(n_2),
\end{align}
\end{widetext}
which shows the separability of the joint probability distribution shown in FIG.~1 c) of the main text.

\section{\label{a:losses}Experimental Losses}
The losses in the experiment were characterized using the Klyshko method \cite{Klyshko1980}. We measured the transmission of the free space part of the experiment (from the source to the fiber coupling) without the fiber beam splitter and with the high efficiency detectors to be $35\% \pm 1\%$. Therefore, these are losses that happen before the interference at the beam splitter. This will be taken into account for the later simulations.

The losses of the two TMDs were also measured individually by measuring the Klyshko efficiency of the PDC source before and after inserting the TMD. The difference between the initial efficiency and the efficiency with the TMD in the setup is the efficiency of the TMD. Both had similar efficiency of $60\%\pm 2\%$.

The detector efficiencies were measured using a laser and calibrated attenuators, where we measured the laser power with a power meter and attenuated it to the single photon level. From this we can get the expected number of photons at the detector and compare it to the measured number of clicks. The resulting detector efficiencies are $94\%\pm 6\%$ and $70\%\pm 6\%$, for the slow and fast detectors, respectively.

The TMD and the detector losses are the losses after the interference at the beam splitter. This will also be taken into account for the simulations.

From this we get the total transmission of the experimental setup of $20\%\pm 6\%$ and $14\% \pm 6\%$, for the high and low efficiency arm respectively. 

\section{\label{a:sim}Simulation}
Here, we describe the theoretical model that we utilize to simulate the measured multi-photon interference effects reported in the main text. The initial state generated by a single-mode type-II PDC source in the high gain regime is a two-mode squeezed state which can be described by applying the operator 

\begin{figure}[hbt]
	\centering
	\includegraphics[width=\linewidth]{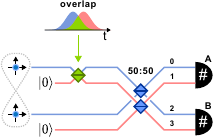}
	\caption{\label{fig:theory_model}Schematic depiction of the used theory model describing the multi-photon interference.}
\end{figure}

\begin{equation}
	S_{ij}(\xi) = \text{exp}\left(\xi^*\hat{a}_i\hat{a}_j - \xi\hat{a}_i^\dag\hat{a}_j^\dag\right)
	\label{eq:squeez}
\end{equation}
to the vacuum. Here, $\xi = r\exp{(i\phi)}$ with $r$ the squeezing parameter and $\phi$ the squeezing angle. $i,j$ label the spatial modes to which the operator is applied. We only consider the case of $\phi=0$, therefore, $\xi = r$. To model the interference of perfectly indistinguishable photons on a beam splitter we could just apply the unitary transformation  
\begin{equation}
	B_{ij}(\theta) = \text{exp}\left(\theta\hat{a}_i\hat{a}_j^\dag - \theta\hat{a}_i^\dag\hat{a}_j\right) 
\end{equation}
to the corresponding spatial modes (with $\theta = \pi/4$ for a 50:50 splitter). However, to model the system including distinguishability we employ a 4-spatial mode model as depicted in FIG.~\ref{fig:theory_model}. The indistinguishable case is modeled via spatial modes 0 and 2 (depicted in blue), where the total TMSVS is interfered on a 50:50 beam splitter. We utilize the additional spatial modes 1 and 3 (in red) to include the contribution of photons which do not interfere due to distinguishability. For introducing the distinguishability, we transfer a part of the state in spatial mode 0 to spatial mode 1 by applying a beam splitter (see FIG.~\ref{fig:theory_model} green beam splitter). The splitting ratio of this beam splitter $\theta_{dis}$ is then mapped to the temporal delay from the experiment via the temporal overlap 
\begin{equation}
	\theta_{dis} = \int_{-\infty}^\infty \text{d}t f_{s}(t) f_i(t)
\end{equation}
between the signal- and idler-fields $f_{s}(t)$ and $f_i(t)$, respectively. The portion of light that is coupled to spatial mode 1 is not interfered with the remaining TMSVS but split by a 50:50 beam splitter and evenly distributed to spatial modes 1 and 3. The total state before detection is then given by 
\begin{equation}
	|\Psi\rangle = B_{13}(\pi/4)B_{02}(\pi/4)B_{01}(\theta_{dis})S_{02}(r)|0\rangle.
\end{equation}
To account for the fact that in the experiment the contributions of distinguishable and indistinguishable photons are detected on the same detectors we model the detection by assuming joint detection of spatial modes 0 and 1, as well as 2 and 3. For this we first calculate the probabilities 
\begin{equation}
	P(\mathbf{n}) = |\langle\mathbf{n}|\Psi\rangle|^2
\end{equation}
of detecting a photon number pattern $\mathbf{n}=(n_0,n_1,n_2,n_3)$ and then combine these via
\begin{equation}
	P(n_A, n_B) = \sum_{\substack{n_B = n_0 + n_1 \\ n_A = n_2 + n_3}} P(\mathbf{n})
\end{equation}
to obtain the probability $P(n_A, n_B)$ of detecting $n_A$ photons in detector A and $n_B$ photons in detector B. 

\section{\label{a:multi_sim}Multimode Simulation}

As a first order approximation to investigate the impact of multimodeness of the PDC source in the simulations presented in section \ref{a:sim} we can include a second TMSVS in the existing simulations. This can be achieved by replacing the vacuum inputs in modes 1 and 3, see Fig. \ref{fig:theory_model}, with a TMSVS. One expects more than a single additional spectral mode in the experimental state \cite{Christ2011}, however, for our nearly single spectral mode source ($K_f=1.3$) two contributing spectral modes are a good first approximation. Performing a rigorous multimode treatment of the interference presented in this work requires a different simulation approach, e.g. following \cite{Bulmer2026}, which is beyond the scope of this work.

\begin{figure*}[hbt]
	\centering
	\includegraphics[width=\linewidth]{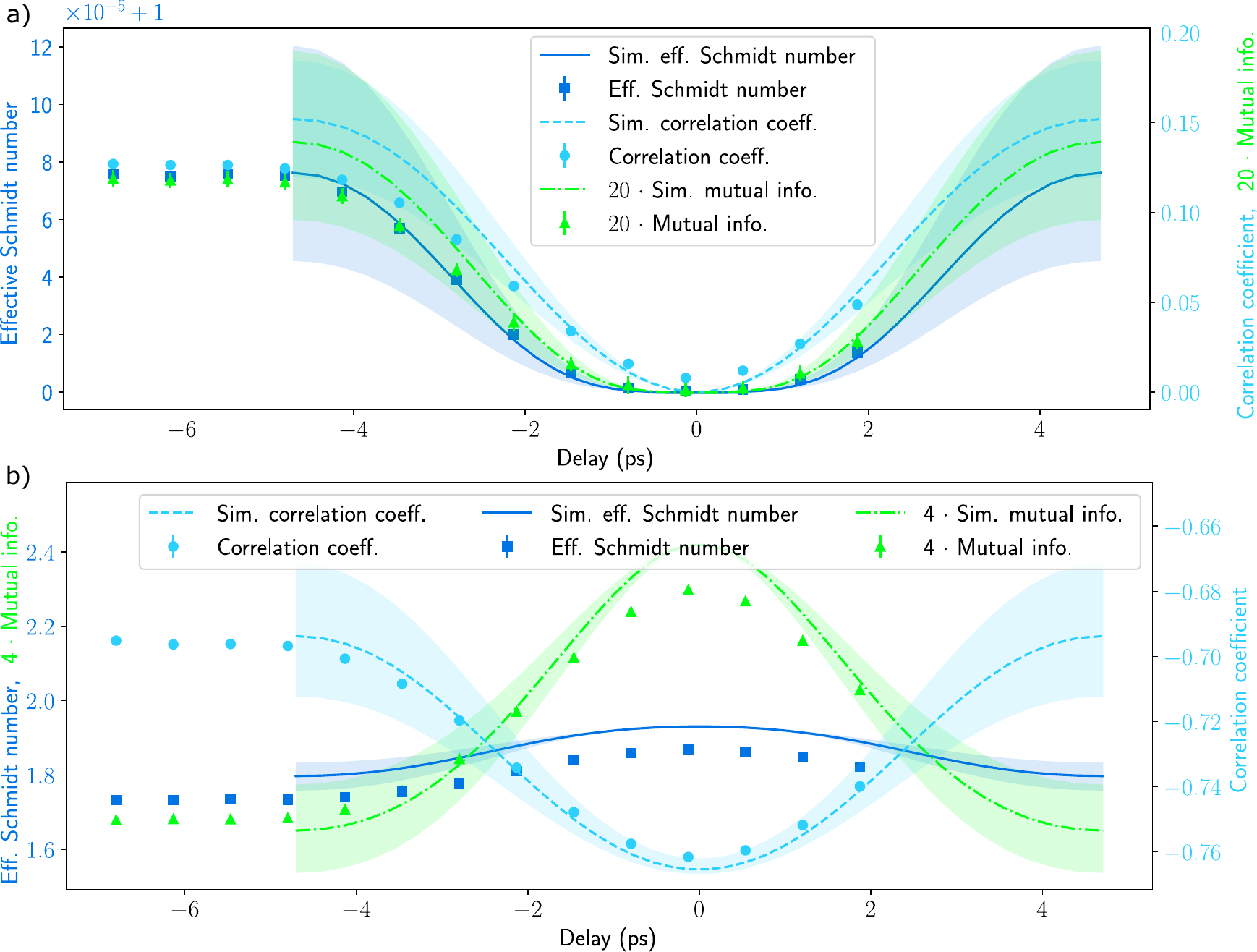}
	\caption{Correlation measurement as a function of the temporal delay between the input pulses. The simulation presented here uses two TMSVSs as inputs with the squeezing parameters $r_1=0.49$ and $r_2=0.037$. a) analysis of the full joint photon number distribution. The experimental data is identical to the one shown in Fig \ref{fig:effk} a) in the main text. b) vacuum component removed from the analysis. The experimental data is identical to the one shown in Fig \ref{fig:trunc} b) in the main text.}
	\label{fig:multimoderesults}
\end{figure*}

The correlation analysis performed to produce Fig. \ref{fig:effk} a) is repeated for the multimode treatment and presented in Fig. \ref{fig:multimoderesults} a). By adjusting the squeezing parameters of both TMSVSs, we are able to find again good agreement between the simulation and the measured data, used the squeezing parameters $r_1=0.49$ and $r_2=0.037$. We see that adding multimode contributions to the simulation improves the overlap to the measured values. However, the squeezing parameters we found result in a Schmidt mode number of around $K_f=1.01$ and therefore are not in agreement with the measured spectral mode number of $K_f=1.3$. If we choose squeezing parameters that match the measured $K_f=1.3$, the simulation predicts more than double the amount of measured correlations for all three correlation measures. Therefore, although the additional mode resulted in an improvement, it did not correspond to the expected Schmidt number $K_f$ which indicated either that modedness is not the issue, or a more complete treatment is required \cite{Christ2011, Bulmer2026}.

The multimode analysis with the vacuum component removed is presented in Fig. \ref{fig:multimoderesults} b). The corresponding single mode simulation is shown in Fig. \ref{fig:trunc} b) in the main text. Comparing both simulations we see a slight improvement of the overlap between the measured and simulated values. The inclusion of a second spectral mode has reduced the offsets for large temporal delay, but did not resolve the offsets at zero temporal delay (see Fig. \ref{fig:multimoderesults} b)).

\begin{figure}[hbt]
	\centering
	\includegraphics[width=\linewidth]{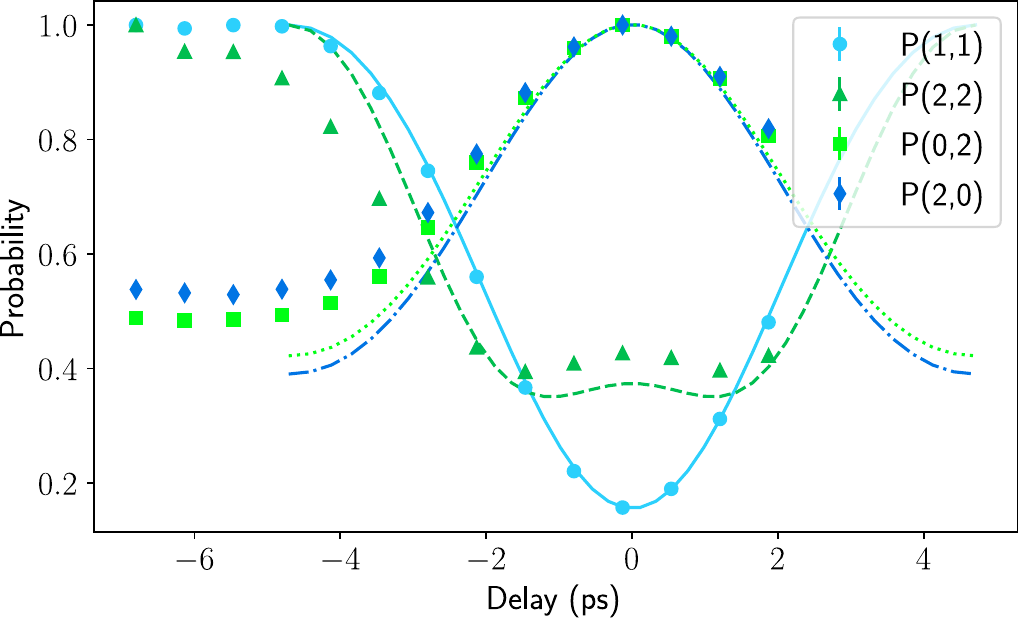}
	\caption{Individual photon number components with the measured data identical to Fig. \ref{fig:pncomp}. Simulations use the squeezing parameters $r_1=0.49$ and $r_2=0.037$.}
	\label{fig:multimodeindividual}
\end{figure}

The individual photon number components of the multimode analysis are shown in Fig. \ref{fig:multimodeindividual}. The single mode results are shown in Fig. \ref{fig:pncomp} in the main text. One can see that the increased probability of the $P(2,2)$ component at a temporal delay of 0 are more pronounced in the multimode case. This matches with the results from \cite{Ferreri2019}, where they have shown that the spectral mode number affects the $P(2,2)$ photon number component. However, we can also see that the multimode simulations do not match the measured data as well for the $P(0,2)$ and $P(2,0)$ components for large time delays (see $-4$~ps in Fig. \ref{fig:multimodeindividual}).

The results from this preliminary multimode simulation show us that multimode contributions do have an effect on the results we obtain. However, we also see that the approach taken here for simulating spectral multimode behavior does not perfectly predict the measured data yet. Further investigations into the effects of spectral multimode contributions are needed, which should include a more sophisticated simulation techniques, like the one presented in \cite{Bulmer2026}, to include a more realistic description of the spectral multimode distribution of a PDC source. Such a treatment would allow us to make a more conclusive statement if the multimode contributions are the main cause of the imperfect predictions of our current simulations or if we have to search for other reasons to explain the deviations.

\section{\label{a:trunc}Conditioning Effect}

Here we show why conditioning on the detection of at least one photon, in the DVQO case, changes two separable SMSVS into an entangled state. To see this, we start with the state of two SMSVS 
\begin{equation}
	\begin{split}    
		\left| r \right\rangle_{a} \left| - r \right\rangle_{b} =  \sum_{n, m=0}^\infty &(-1)^m \frac{\sqrt{(2n)!(2m)!}}{n!m!}  \\
		&\cdot\left(\frac{\tanh r}{2}\right)^{n+m} \left| 2n \right\rangle_a \left| 2m \right\rangle_b,
	\end{split}
\end{equation}
where we neglect the normalization. The key point in our analysis is that this state will be modified if we have further information about it and that this information can leave the state entangled. There are two distinct types of information that we may have: \textit{uncorrelated} and \textit{correlated} information. It is fundamentally the difference between these that leaves the modified state separable or entangled. Let us consider each in turn.

\subsection{Uncorrelated Information}
Uncorrelated information tells us something about the state of one of the modes or of them both but does not reveal any correlated or joint information. A simple example, appropriate for the small squeezing limit, is the statement that both modes have either two photons or are in their vacuum state. With this information, we must modify the state to remove all components in which either mode has four or more photons, with the result
\begin{equation}
	\begin{split}
		\left| r \right\rangle_{a} \left| - r \right\rangle_{b} \to &\left| 0 \right\rangle_{a} \left| 0 \right\rangle_{b}\\ & + \frac{\tanh r}{\sqrt{2}} (\left| 2 \right\rangle_{a} \left| 0 \right\rangle_{b} - \left| 0 \right\rangle_{a} \left| 2 \right\rangle_{b}) \\&- \frac{\tanh^2 r}{2}\left| 2 \right\rangle_{a} \left| 2 \right\rangle_{b}.
	\end{split}
\end{equation}
This state is clearly separable
\begin{equation}
	\left(\left| 0 \right\rangle_{a} + \frac{\tanh r}{\sqrt{2}}\left| 2 \right\rangle_{a}\right) \left(\left| 0 \right\rangle_{b} - \frac{\tanh r}{\sqrt{2}}\left| 2 \right\rangle_{b}\right)
\end{equation}
and this is a consequence of the fact that the information added has told us nothing about the correlations
between the modes.

\subsection{Correlated Information}
It is a different story if the additional information is correlated, i.e. if it tells us something about the joint properties of the two modes. The simplest (and also most natural) example is if we know (or require) that there are at least some photons present. Equivalently, we can state that the two-mode state is then orthogonal to the two-mode vacuum state. It is essential to appreciate that this is indeed a correlated piece of information: we are not stating that either mode is in the vacuum state, but rather that only one of the two modes can be in its vacuum state. If we again ignore normalization then the state is modified as
\begin{equation}
	\left| r \right\rangle_{a} \left| - r \right\rangle_{b} \to \left| r \right\rangle_{a} \left| - r \right\rangle_{b} - \left| 0 \right\rangle_{a} \left| 0 \right\rangle_{b}.
\end{equation}
The simplest way to see that this is indeed entangled is to see what happens to the state of mode $b$ when we measure the photon number in mode $a$. There are two distinct cases: In the first, if the measurement on mode $a$ reveals the presence of photons, then mode $b$ is left in the state $\left| - r \right\rangle_{b}$. If, however, the measurement shows mode $a$ to be in its vacuum state the state of mode $b$ is modified to $\left| - r \right\rangle_{b} - \left| 0 \right\rangle_{b}$, i.e. the squeezed vacuum state without its vacuum component. The mere fact that we started with a pure two-mode state and that different measurement outcomes on mode $a$ leave mode $b$ in different states suffices to ensure that the initially separable two-mode state becomes entangled when modified by \textit{correlated} information.

We note that the truncation of the state by the removal of amplitudes corresponding to more than two photons (in each mode) does not, alone, suffice to produce an entangled state. To produce an entangled state we need this truncation to be in the form of correlated information as, for example, the statement that the two modes, between them, have precisely (or at most) two photons.

\bibliographystyle{quantum}
\bibliography{bib_file}

\end{document}